\pdfoutput=1
\documentclass[11pt,a4paper]{article}
\usepackage{geometry}
\usepackage{amsfonts}
\usepackage{amssymb}
\usepackage[centertags]{amsmath}
\usepackage{graphicx}
\usepackage{booktabs}
\usepackage{theorem}
\usepackage{array}
\usepackage{ulem}
\usepackage[small,bf]{caption}
\usepackage{cite}
\usepackage{color}
\usepackage{colordvi}
\usepackage[colorlinks,citecolor=blue]{hyperref}

\newcommand{\FMDG}[1]{\ensuremath{\vcenter{\hbox{\includegraphics{#1}}}}}

\usepackage{subcaption}

\allowdisplaybreaks[1]

\numberwithin{equation}{section}

\newcommand{\vev}[1]{\ensuremath{\left\langle#1\right\rangle}}

\DeclareMathOperator{\ci}{\text{i}}

\newcommand{\Secref}[1]{Sec.~\ref{#1}}
\newcommand{\Eqref}[1]{eq.~\eqref{#1}}

\makeatletter

\newcommand{\Rmnum}[1]{\expandafter\@slowromancap\romannumeral #1@}
\makeatother

\begin{document}

\begin{titlepage}

\vspace*{-15mm}
\begin{flushright}
MPP-2013-173\\
SISSA 27/2013/FISI
\end{flushright}
\vspace*{0.7cm}

\begin{center}
{
\bf\LARGE
Solving the Strong CP Problem with Discrete Symmetries and the Right Unitarity Triangle
}
\\[8mm]
Stefan~Antusch$^{(a)}$
\footnote{E-mail: \texttt{stefan.antusch@unibas.ch}},
Martin~Holthausen$^{(b)}$
\footnote{E-mail: \texttt{martin.holthausen@mpi-hd.mpg.de}},
Michael~A.~Schmidt$^{(c)}$
\footnote{E-mail: \texttt{michael.schmidt@unimelb.edu.au}},
Martin~Spinrath$^{(d)}$
\footnote{E-mail: \texttt{spinrath@sissa.it}},
\\[1mm]
\end{center}
\vspace*{0.50cm}
\centerline{$^{(a)}$ \it
 Department of Physics, University of Basel,}
\centerline{\it
Klingelbergstr.~82, CH-4056 Basel, Switzerland}
\vspace*{0.2cm}
\centerline{$^{(a)}$ \it
Max-Planck-Institut f\"ur Physik (Werner-Heisenberg-Institut),}
\centerline{\it
F\"ohringer Ring 6, D-80805 M\"unchen, Germany}
\vspace*{0.2cm}
\centerline{$^{(b)}$ \it
Max-Planck-Institut f\"ur Kernphysik,}
\centerline{\it
Saupfercheckweg 1, D-69117 Heidelberg, Germany }
\vspace*{0.2cm}
\centerline{$^{(c)}$ \it
ARC Centre of Excellence for Particle Physics at the Terascale,}
\centerline{\it
School of Physics, The University of Melbourne, Victoria 3010, Australia}
\vspace*{0.2cm}
\centerline{$^{(d)}$ \it
SISSA/ISAS and INFN,}
\centerline{\it
Via Bonomea 265, I-34136 Trieste, Italy }
\vspace*{1.20cm}

\begin{abstract}
\noindent
We present a solution to the strong CP problem based
on spontaneous CP violation and discrete family
symmetries. The model predicts in a 
natural way the almost right-angled quark unitarity
triangle angle ($\alpha \simeq 90^\circ$) by making the entries of the quark mass
matrices either real or imaginary. By this choice
the determinants of the mass matrices are rendered real
and hence the strong CP phase vanishes.
We present a toy model for the quark
sector that demonstrates the viability of our approach.
\end{abstract}

\end{titlepage}

\setcounter{footnote}{0}

\section{Introduction}
\label{sec:Introduction}

From a multitude of experimental observations, quantum chromodynamics (QCD) has emerged as the well-established theory of strong interactions. However, the smallness of CP violation in strong interactions has been a puzzle in particle physics since the 1970s when it was realised that the QCD Lagrangian violates CP due to instanton effects \cite{Belavin:1975fg, 'tHooft:1976up}. The CP violation in strong interactions is described by the strong CP  phase
\begin{equation} \label{eq:DefThetaBar}
 \bar \theta = \theta + \arg \det (M_u M_d) \;, 
\end{equation}
where $\theta$ is the coefficient of $\alpha_s/(8\pi) \tilde G_{\mu \nu} G^{\mu \nu}$, $G_{\mu\nu}$ is the field strength tensor of QCD, $\tilde G_{\mu\nu}$ its dual, and $\arg\det (M_u M_d)$ is the contribution from the quark masses. While $\theta$ and $ \arg \det (M_u M_d)$ are transformed into each other via a chiral transformation, the combination $\bar \theta$ stays invariant. The most stringent limits originate from experimental bounds on the electric dipole moment of the neutron and result in $\bar \theta \lesssim 10^{-11}$~\cite{Beringer:1900zz, Burghoff:2011xk}, which is much smaller than the Jarlskog invariant, $J=\left(2.96^{+0.20}_{-0.16}\right)\times10^{-5}$~\cite{Beringer:1900zz}.
Therefore, the strong CP problem is the question why the two contributions to $\bar \theta$ sum up to such a small number.

There are three main ideas to explain the smallness
of strong CP violation. 
The first and simplest solution is that one of the quarks is massless
\cite{'tHooft:1976up}. In this case the strong CP phase $\bar\theta$ is unphysical, since it can be absorbed in the massless quark by a phase transformation.
However, recent data strongly suggests that
all quarks are massive \cite{Beringer:1900zz}.

The second very popular solution is the axion solution
\cite{Peccei:1977hh} where $\bar \theta$ is promoted to
a dynamical degree of freedom which is set to small values
by a potential. This solution is very elegant but albeit
there have been extensive searches for axions there have
been no experimental hints for their existence so far
\cite{Beringer:1900zz}.

The third approach solves the strong CP problem by breaking parity (or CP) spontaneously. 
As the topological term $\alpha_s/(8\pi) \tilde G_{\mu \nu} G^{\mu \nu}$ violates parity (as well as CP), there are two possibilities to forbid it by either imposing parity and/or CP, which we will briefly discuss in the following.

(i) Left-right symmetric theories naturally conserve parity and therefore predict $\bar\theta=0$. This has been pointed out in Ref.~\cite{Mohapatra:1978fy} and further developed in Ref. \cite{Barr:1991qx}. Although the Yukawa couplings are Hermitian, a solution to the strong CP problem requires that the breaking of parity does not introduce a complex phase in the mass matrices via a complex phase of a vacuum expectation value (vev). However, there are several viable models in the literature. See, e.g., \cite{Barr:1991qx,Kuchimanchi:2010xs} for non-supersymmetric models and Ref.~\cite{Mohapatra:1996vg} for a supersymmetric (SUSY) model.

(ii) 
Promoting CP to
a fundamental symmetry of the Lagrangian sets $\bar\theta = 0$. In order to explain the CP violation in weak processes, CP has to be broken spontaneously~\cite{Lee:1973iz} in such a way that $\arg \det (M_u M_d)$ stays sufficiently
small, while the CP violation in weak interactions is large.
The most popular class of models accommodating this are the Nelson-Barr models \cite{Nelson:1983zb,Barr:1984qx}. See, for instance, Ref.~\cite{Bento:1991ez} for a minimal implementation as well as Ref.~\cite{Barr:1988wk} for an implementation within SUSY. 
In supersymmetry the smallness of the strong CP phase is further protected by the non-renormalization theorems~\cite{Ellis:1982tk}. This has been used in the SUSY model of Ref.~\cite{Hiller:2001qg}, where a large CKM phase is generated by renormalization group running.
Obviously, it is also possible to invoke parity as well as CP conservation to address the strong CP problem, which has been used in an extra-dimensional model with split fermion profiles in Ref.~\cite{Harnik:2004su}.
Recently, Fong and Nardi~\cite{Fong:2013sba} showed that by promoting the Yukawa couplings to spurions of the maximal $SU(3)^3$ flavour symmetry, the spurion potential results in a real determinant and therefore a solution of the strong CP problem.

Nevertheless, the class of models proposed here is based as well on spontaneous CP violation, but different to the previously mentioned class of models by using a specific texture of quark mass matrices. As we will discuss in the next section, where we outline our strategy, our class of models
is based on a sum rule for the phases in the 
CKM matrix \cite{Antusch:2009hq} suggesting a simple
structure for quark mass matrices with either real or purely imaginary elements \cite{rightangledothers}.  See, for example, \cite{Barr:1996wx,Barr:1997ph,Masiero:1998yi,Glashow:2001yz,Chang:2003uq} for an incomplete list of models with different mass matrix textures. This simple
structure finds a natural realisation in flavour models based on non-Abelian discrete family symmetries where the CP
symmetry is spontaneously broken with a method dubbed discrete vacuum alignment \cite{Antusch:2011sx}. This method was previously used in various flavour models \cite{Meroni:2012ty, Antusch:2013wn,Antusch:2013kna,Antusch:2013tta}.

Before presenting the details, let us highlight the main ingredients and achievements: (i) We use non-Abelian discrete family symmetries, which allow an implementation of our solution to the strong CP problem in successful models of flavour. This is the first paper addressing the strong CP problem with non-Abelian discrete family symmetries to our knowledge. (ii) It uses the discrete vacuum alignment mechanism~\cite{Antusch:2011sx}, which also predicts the CP phases of the vevs up to a discrete choice. (iii) The implementation in a SUSY theory protects the smallness of the strong CP phase by the non-renormalization theorems~\cite{Ellis:1982tk}. (iv) The main prediction of our toy model, besides the smallness of the strong CP phase, is the correct prediction of the CKM phase, while all other flavour parameters in the quark sector can be accommodated.

Our paper is organised as follows. After presenting the general strategy for our solution to the strong CP problem in \Secref{sec:Strategy}, we discuss several contributions to the strong CP phase $\bar\theta$ from SUSY breaking in \Secref{sec:SUSYbreaking}. In \Secref{sec:Model}, we discuss a model of quark flavour, which implements our solution to the strong CP problem and compare it to models based on spontaneous breaking of CP in \Secref{sec:NelsonBarr}. Finally, we conclude in \Secref{sec:conclusions}.

\section{The strategy}
\label{sec:Strategy}

In this paper we present a solution for the strong CP problem based on
spontaneous CP violation. If CP is a fundamental symmetry of the Lagrangian, the strong CP phase $\bar\theta$ will vanish. However, in order to explain CP violation in weak interactions, the spontaneous breaking of CP has to explain the large value of the CKM phase, while the strong CP phase $\bar \theta$ vanishes or is tiny enough to be in agreement with experimental data. 

In other words we have to look for a texture with $\arg \det (M_u M_d) = 0$
and a realistic value for the CKM phase. Furthermore, if we do not want to
assume cancellations between the phases in the up and the down sector
$\det M_u$ and $\det M_d$ should be real and positive by itself already.

One possible choice is, for instance, that $M_u$ is completely
real and has negligible 1-3 mixing, and that 
\begin{equation} \label{eq:MdStructure}
 M_d = \begin{pmatrix}
	0 & * & 0 \\
	* & \ci * & * \\
	0 & 0 & *
       \end{pmatrix} \;,
\end{equation}
where $'*'$ are some real entries. The only non-trivial complex phase appears
in the 2-2 element of $M_d$ and the determinants of both mass matrices are
real. Note that for simplicity we assume here both signs of the determinants
to be positive.

If this structure of the mass matrices can be realised from the spontaneous
breaking of CP we would indeed have a solution for the strong CP problem.
And furthermore this very simple structure can also correctly reproduce
the right quark unitarity triangle, as it was demonstrated in \cite{Antusch:2009hq}, 
since it satisfies the phase sum rule
\begin{equation}
\alpha \approx \delta_{12}^d - \delta_{12}^u \approx 90^\circ  \;,
\end{equation}
where $\alpha$ is the angle of the CKM unitarity triangle
measured to be close to $90^\circ$ \cite{Beringer:1900zz} and
$\delta_{12}^{d/u}$ are the phases of the complex 1-2 mixing angles
diagonalizing the quark mass matrices (for the conventions used, see
\cite{Antusch:2009hq}). Now any model, which generates such a structure
can do the trick, but the question is, whether such models exist.

Before we will discuss a toy model in \Secref{sec:Model} we outline
how this could be achieved in the context of discrete family symmetries
which have gained a lot of attention for describing the mixing in the
lepton sector.

Suppose we have a family symmetry $G_F$ with triplet representations
(we will use later on $A_4$, but $S_4$, $T'$, $\Delta(27)$, etc.\ 
would work equally well). See Ref.~\cite{King:2013eh} for a recent review on flavour symmetries. Then we assume the right-handed down-type
quarks to transform as triplets under $G_F$ while all
other quarks are singlets. The rows of $M_d$ are then proportional
to the vacuum expectation values (vevs) of family symmetry breaking
Higgs fields, the so-called flavon fields, which are triplets under $G_F$.
$M_u$ is generated by vevs of singlet flavon fields.

Introducing four flavon triplets with the following alignments in flavour
space
\begin{equation}\label{eq:vevs}
 \langle \phi_1 \rangle \sim \begin{pmatrix} 1 \\ 0 \\ 0 \end{pmatrix} \;,\quad
 \langle \phi_2 \rangle \sim \begin{pmatrix} 0 \\ 1 \\ 0 \end{pmatrix} \;,\quad
 \langle \phi_3 \rangle \sim \begin{pmatrix} 0 \\ 0 \\ 1 \end{pmatrix} \;,\quad
 \langle \tilde \phi_2 \rangle \sim \ci \begin{pmatrix} 0 \\ 1 \\ 0 \end{pmatrix} \;,\quad
\end{equation}
which can be achieved by standard vacuum alignment techniques
we can reproduce the desired structure for $M_d$. Note that we have
explicitly written out any complex phases (we assume $\langle \phi_i \rangle$,
$i=1,2,3$, to be real while only $\langle \tilde \phi_2 \rangle$ is purely imaginary).

Indeed it is not quite trivial to fix the phases of these vevs. The method described in \cite{Antusch:2011sx}, which we want to sketch here for a singlet flavon field $\xi$, is one possibility to achieve it. Suppose $\xi$ is charged
under a discrete $Z_n$ symmetry and apart from that neutral then we can write down
a superpotential for $\xi$
\begin{equation}
\mathcal{W} = P \left( \frac{\xi^n}{\Lambda^{n-2}} \mp M^2 \right) \;,
\label{eq:W-strucutre}
\end{equation}
where $P$ is a total singlet and $M$ and $\Lambda$ mass scales.
We have dropped prefactors for brevity and since we assume fundamental CP symmetry these prefactors and the mass scales
are real.\footnote{Note that we use the generalised CP transformation, which is trivial with respect to $A_4$. It agrees with the ordinary CP transformation for real representations of $A_4$. See \cite{Feruglio:2012cw,Holthausen:2012dk} for a recent discussion of generalised CP in the context of non-Abelian discrete symmetries.} Without loss of generality we assume the prefactors and $M^2$ to be positive: a possible relative sign can be absorbed by changing $\mp M^2$ to $\pm M^2$.
From the potential for $\xi$,
\begin{equation} \label{eq:flavonpotentialZn}
|F_P|^2 =  \left| \frac{\xi^n}{\Lambda^{n-2}} \mp M^2 \right|^2 .
\end{equation}
Since $|F_P| = 0$  the vev of $\xi$ has to satisfy
\begin{equation}
\langle \xi^n \rangle = \pm \,\Lambda^{n-2}  M^2\;.
\end{equation}
and hence
\begin{equation}\label{eq:phaseswithZn}
\arg(\langle \xi \rangle) =   \left\{ \begin{array}{ll}
\frac{2 \pi}{n}q \;,\quad q = 1, \dots , n & \mbox{\vphantom{$\frac{f}{f}$} for ``$-$'' in Eq.~(\ref{eq:flavonpotentialZn}),}\\
\frac{2 \pi}{n} q +\frac{\pi}{n} \;,\quad q = 1, \dots , n & \mbox{\vphantom{$\frac{f}{f}$} for ``$+$'' in Eq.~(\ref{eq:flavonpotentialZn}).}
\end{array}
\right.
\end{equation} 
Here the phases of the vevs do not depend on potential parameters, a situation which has been dubbed 'calculable phases' in the literature\cite{Branco:1983tn}. In Ref.~\cite{Holthausen:2012dk} this phenomenon was interpreted as the result of an accidental CP symmetry of the potential. The same discussion applies here. For a real coupling, the potential (\ref{eq:W-strucutre}) is invariant under the CP transformation $\xi\rightarrow z \xi^*$ with $z^n=1$. This generalized CP transformation emerges as an accidental symmetry of the potential but will be explicitly broken elsewhere (e.g. in the couplings to the matter sector). If the whole Lagrangian was invariant under this CP transformation, then there would be no CP violation in physical observables. We furthermore note that it is necessary to break CP with two different fields~\cite{Haber:2012np}, such that it is impossible to define a CP transformation, which is left invariant by the vevs, and CP is spontaneously broken. The interplay between the different flavon fields will ensure that CP is broken in our model, as will be discussed in \Secref{sec:Model:symmetries}.

Using these ingredients we will present in \Secref{sec:Model}
a consistent flavour model with spontaneous CP violation for the
quark sector which resembles a real $M_u$ and the structure of 
$M_d$ from eq.~\eqref{eq:MdStructure}.

Due to the stringent constraints on $\bar \theta$, special care needs to be taken with corrections to this parameter. The most important corrections are:
\begin{itemize}
\item Higher dimensional contributions to the superpotential that would spoil the structure of the mass matrices.
\item Corrections which are induced from SUSY breaking terms.
\end{itemize}
In the following we will discuss these corrections. The first point will be addressed by
introducing shaping symmetries fixing the phases of the flavon vevs as well as by specifying the messenger sector which gives us full control over all higher order operators. Even a small higher order contribution $\delta M$ would contribute to $\bar \theta$ as
\begin{align}
\delta \bar \theta \approx \arg \det (\delta M \, M^{-1}) \;,
\end{align}
which has to be smaller than $10^{-11}$.
The same applies to corrections coming from the
SUSY sector of the theory which we discuss in the upcoming section.

\section{Corrections from SUSY breaking}
\label{sec:SUSYbreaking}
There are two important consequences when a solution to the strong CP problem is applied to a SUSY model: On the one hand, as long as SUSY is unbroken, non-renormalisation theorems guarantee that $\bar \theta$ will not be generated radiatively at any loop order. On the other hand, the SUSY breaking sector can in principle also introduce new sources of CP violation, which can then have an impact on $\bar \theta$ (see e.g.~\cite{Dine:1993qm, Barr:1997ph, Hiller:2001qg}). Before we turn to the construction of an example flavour model where our strategy of  \Secref{sec:Strategy} is realised, let us therefore discuss the possible corrections to $\bar \theta$ from SUSY breaking. We note that although our general strategy applies also to non-SUSY models, our example model will be formulated in a SUSY framework and also our method to fix the phases of the flavon vevs, and thus the phases of the mass matrix entries, relies on SUSY. 

\begin{figure}\centering
\begin{subfigure}{5cm}
\FMDG{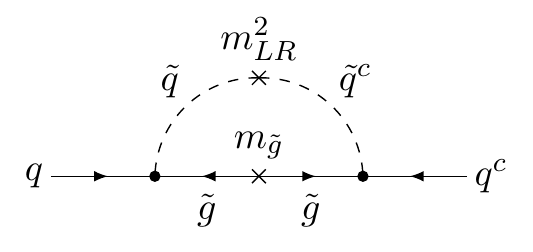}
\caption{Contribution to quark mass.}
\end{subfigure}
\hspace{1cm}
\begin{subfigure}{5cm}
\FMDG{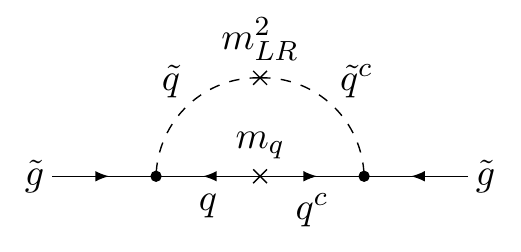}
\caption{Contribution to gluino mass.}
\end{subfigure}
\caption{Diagrams contributing to $\bar \theta$ in a theory with broken SUSY.
\label{fig:SUSYTheta}}
\end{figure}

To illustrate the possible effects of SUSY breaking on $\bar \theta$, we start by noting that with a general complex gluino mass parameter $m_{\tilde g}$, $\bar \theta$ would get an additional contribution of the form  
$\delta \bar \theta = 3 \arg (m_{\tilde g})$.  
Furthermore, there is a contribution from SUSY loop corrections to the quark
mass matrices and the gluino mass, as shown in Fig.~\ref{fig:SUSYTheta}. In general
these corrections depend on plenty of SUSY breaking
parameters, for instance, on the trilinear couplings. Explicit formulae can be found, e.g.\ in \cite{Pierce:1996zz}. Also in the MSSM with complex parameters the Higgs vevs $v_u$ and $v_d$ can become complex and may in principle introduce additional CP violating phases. However, fortunately, many of these potential sources of corrections to $\bar \theta$ are safely under control. Due to our assumption that the fundamental theory conserves CP, one could easily imagine that the SUSY breaking potential by itself does not introduce CP breaking. 
Then, parameters like $m_{\tilde g}$ and the $\mu$ parameter are real and various potential corrections to $\bar \theta$ vanish.

In this case, the discussion of corrections from SUSY breaking boils down to the question of how well the conventional SUSY flavour and CP problem is solved. In this paper, we will not construct an explicit SUSY breaking sector, but rather refer to the discussion on this issue which already exists in the literature, and add some remarks on the connection to non-Abelian family symmetries: To start with, it has been discussed, e.g.\ in \cite{Hiller:2001qg}, that anomaly mediation or gauge mediation can in principle provide useful frameworks. In the context of flavour models with Abelian symmetries, a solution by ``flavour alignment'' has been suggested by Nir and Seiberg in Ref. \cite{Nir:1993mx} and a discussion in the context of solutions to the strong CP problem can be found in \cite{Barr:1996wx}. 

On the other hand, flavour models with non-Abelian family symmetries by themselves can provide promising frameworks for solving the SUSY flavour problem (see, e.g. \cite{Abel:2001cv, Ross-Vives, Ross:2004qn,Antusch:2007re,Olive:2008vv, su3edms,Feruglio:2009hu,Calibbi:2010rf,Lalak:2010bk}). With the three families of matter fields embedded into triplet representations of a non-Abelian family symmetry group $G_F$, the soft terms are universal before family symmetry breaking, and non-universalities only get induced after spontaneous $G_F$ breaking. This allows to control the flavour (and CP) structure of the SUSY breaking terms in explicit ``SUSY-flavour'' models. 

When constructing a ``SUSY-flavour'' model of this type, care has to be taken of the contributions to the soft terms from flavon $F$-terms \cite{Antusch:2008jf,Feruglio:2009iu}. These contributions are especially relevant, because if CP symmetry is broken by the flavon vevs, their $F$-terms can in principle generate a CP violating non-universality in the $A$-terms and might thereby introduce a sizeable contribution to $\bar \theta$, e.g.\ via the diagrams in Fig.~\ref{fig:SUSYTheta}. In a supergravity scenario with sequestered K\"ahler and superpotential (as, e.g.\ in \cite{Antusch:2011sq}), universalities in the $A$-terms would only stem from the flavon $F$-terms, so controlling them is crucial. In \cite{Ross-Vives} it has been argued that their size is typically of the order $m_{3/2} \langle \phi \rangle$, where $m_{3/2}$ is the gravitino mass and  $\langle \phi \rangle$ is a flavon vev, which could easily spoil the solution to the strong CP problem.

However, it has been shown in \cite{Antusch:2008jf} that the flavon $F$-terms are strongly suppressed for flavon superpotentials with driving fields, as we are going to use in this paper, by powers of $m_{3/2}/\Lambda$, with $\Lambda$ being the messenger scale of the flavour model.\footnote{Furthermore, in supergravity the flavon vevs can provide an additional contribution to the flavour structure via higher-dimensional operators in the K\"ahler potential, leading to corrections via canonical normalisation (see, e.g.~\cite{CN}). The size of these corrections depends on the details of the messenger sector of the model. However, in any case, canonical normalisation cannot induce a contribution to the $\bar \theta$ term \cite{Ellis:1982tk}.}
Such a suppression would render them harmless to the solution of the strong CP problem. 
 Without going into further details, we conclude that models of the class we propose in this paper, amended by a suitable SUSY breaking sector, have the potential to be safe from dangerously large corrections to $\bar \theta$.

\section{The model}
\label{sec:Model}
In this section we aim to flesh out the preceding discussion by constructing an explicit model which conforms to the general strategy discussed in Section~\ref{sec:Strategy} and assume a SUSY breaking sector along the lines of the discussion in the preceding section. While the model presented here only discusses the quark sector, it employs model building techniques that were primarily used to describe the lepton sector. There, it has been long known to be useful to assign the left-handed lepton doublets to three-dimensional irreducible representations of some non-Abelian flavour group. After the discrete flavour symmetry is broken in carefully chosen directions in flavour space (usually corresponding to invariant subgroups of named groups), mixing angles can be predicted (see, for example, \cite{King:2013eh, deAdelhartToorop:2011re} for an overview).

\subsection{Symmetries and model setup}
\label{sec:Model:symmetries}
For our model we will use the group $A_4=\langle S,T\vert S^2=T^3=(ST)^3=E\rangle $, the smallest discrete group with a three-dimensional irreducible representation
\begin{align}
\rho(S)&=\left(\begin{array}{ccc}
1&0&0\\
0&-1&0\\
0&0&-1
\end{array}\right),
& 
\rho(T)&=\left(\begin{array}{ccc}
0&1&0\\
0&0&1\\
1&0&0
\end{array}\right) ,
& 
\label{eq:Ma-basis}
\end{align}
in the basis which is sometimes called the Ma-Rajasekaran\cite{Ma:2001dn} basis. We follow an approach to flavour model building\footnote{For a review of flavour model building of this general type, the reader is referred to \cite{King:2013eh,Ishimori:2012zz}.} which employs three scalar fields (flavons) $\phi_1$, $\phi_2$ and $\phi_3$ transforming as triplets and breaking the group $A_4$ via the vevs 
 \begin{equation}
  \langle \phi_1 \rangle \sim \begin{pmatrix} 1 \\ 0 \\ 0 \end{pmatrix} \;,\quad
  \langle \phi_2 \rangle \sim \begin{pmatrix} 0 \\ 1 \\ 0 \end{pmatrix} \;,\quad
  \langle \phi_3 \rangle \sim \begin{pmatrix} 0 \\ 0 \\ 1 \end{pmatrix} \;,
 \end{equation}
down to the subgroups generated by $S$, $T^2ST$ and $TST^2$, respectively. $A_4$ is frequently used in flavour model building, since it allows to readily realise the observed large lepton mixing (which we will not consider here) and since it is the smallest discrete group with triplet representations.

\begin{table}
\centering
\begin{tabular}{c ccccc cccc}
\toprule
 & $G_{\text{SM}}$ & $A_4$ & $U(1)_R$ & $Z_2$ & $Z_4$ & $Z_4$ & $Z_4$ & $Z_4$ & $Z_4$ \\ 
 \midrule 
$\phi_{1}$ & $(\mathbf{1},\mathbf{1},0)$ & $\mathbf{3}$  & 0 & 1 & 1 & 3 & 3 & 1 & 2\\
$\phi_{2}$ & $(\mathbf{1},\mathbf{1},0)$ & $\mathbf{3}$  & 0 & 1 & 3 & 0 & 3 & 2 & 2\\
$\phi_{3}$ & $(\mathbf{1},\mathbf{1},0)$ & $\mathbf{3}$  & 0 & 1 & 2 & 3 & 0 & 2 & 2\\
$\tilde \phi_{2}$  & $(\mathbf{1},\mathbf{1},0)$ & $\mathbf{3}$  & 0 & 0 & 1 & 0 & 0 & 2 & 2\\
\midrule
$\xi_{d}$ & $(\mathbf{1},\mathbf{1},0)$ & $\mathbf{1}$  & 0 & 0 & 0 & 3 & 0 & 0 & 1\\
$\xi_{s}$ & $(\mathbf{1},\mathbf{1},0)$ & $\mathbf{1}$  & 0 & 1 & 0 & 2 & 3 & 3 & 1\\
$\xi_{u}$ & $(\mathbf{1},\mathbf{1},0)$ & $\mathbf{1}$  & 0 & 0 & 2 & 1 & 0 & 1 & 0\\
$\xi_{c}$ & $(\mathbf{1},\mathbf{1},0)$ & $\mathbf{1}$  & 0 & 0 & 2 & 2 & 2 & 2 & 2\\
$\xi_{t}$ & $(\mathbf{1},\mathbf{1},0)$ & $\mathbf{1}$  & 0 & 0 & 3 & 3 & 3 & 3 & 1\\
\bottomrule
\end{tabular}
\caption{\label{Tab:FlavonFields}
The flavon fields in our model and their quantum numbers. $G_{\text{SM}}$ is the Standard Model gauge group $SU(3) \times SU(2) \times U(1)_Y$.}
\end{table}

We further impose a fundamental CP symmetry on the Lagrangian. As was recently discussed in Ref. \cite{Holthausen:2012dk}, care has to be taken when defining a CP transformation in the context of discrete flavour symmetry groups. A non-trivial CP transformation has to act on the group of internal symmetries as an outer automorphism. For the case of $A_4$, it was shown that if the theory contains a non-trivial singlet field, CP has to act in a non-trivial way in flavour space (for the basis given above). Since we do not introduce such fields here, we are free to define our fundamental CP transformation as every field going to its complex conjugate
\begin{align}
\mathrm{CP:} \qquad \varphi(t,\vec{x}) \rightarrow \varphi^*(t,-\vec{x}).
\end{align}
 Note that this transformation corresponds to an inner automorphism of $A_4$ and an outer automorphism of the standard model gauge group $G_{\text{SM}}$ and is thus able to give non-trivial predictions.

We build our model such that this CP transformation is broken only by the vev of a single triplet field 
\begin{align}
 \langle \tilde \phi_2 \rangle \sim \ci \begin{pmatrix} 0 \\ 1 \\ 0 \end{pmatrix}.
 \label{eq:CPvioaltingvev}
\end{align}
Note that this vev leaves the CP transformation invariant, where $ \tilde \phi_2$ and $ \xi_s$ pick up a sign,
\begin{align}
\mathrm{CP^\prime}: \tilde \phi_2\rightarrow -\tilde\phi_2^* \qquad \xi_s \rightarrow -\xi_s^*\;,
\end{align}
because the combination $\tilde \phi_2 \xi_s$ couples to the matter sector.
This CP transformation is a symmetry of the Lagrangian, but it is spontaneously broken by the real vev of $\xi_s$.
If $\xi_s$ had an imaginary vev, this form of CP would force physical CP observables to vanish, in accord with the discussion in \cite{Holthausen:2012dk}. It is also well known that one needs at least two fields to obtain spontaneous CP violation, see, for instance, \cite{Haber:2012np}. 

To arrange for this vev configuration to be dynamically realised along the lines outlined in Section \ref{sec:Strategy}, additional symmetries and fields have to be introduced. We will relegate the details of the vev alignment mechanism to Section \ref{sec:Model:Flavon}, and first discuss the matter sector of the model.

\subsection{Matter sector}
\label{sec:Model:Matter}

\begin{table}
\centering
\begin{tabular}{c cccccc ccccc}
\toprule
 &  $Q_1$ & $Q_2$ & $Q_3$ & $\bar u_1$ & $\bar u_2$ & $\bar u_3$ & $\bar d$ & $H_u$ & $H_d$ \\ \midrule 
$SU(3)$ & $\mathbf{3}$ & $\mathbf{3}$ & $\mathbf{3}$ & $\bar{\mathbf{3}}$ & $\bar{\mathbf{3}}$ & $\bar{\mathbf{3}}$ & $\bar{\mathbf{3}}$ & $\mathbf{1}$ & $\mathbf{1}$ \\ 
$SU(2)$ & $\mathbf{2}$ & $\mathbf{2}$ & $\mathbf{2}$ & $\mathbf{1}$ & $\mathbf{1}$ & $\mathbf{1}$ & $\mathbf{1}$ & $\mathbf{2}$ & $\mathbf{2}$  \\ 
$U(1)_Y$ & $ \frac{1}{3}$ & $ \frac{1}{3}$ & $ \frac{1}{3}$ & $- \frac{4}{3}$ & $- \frac{4}{3}$ & $- \frac{4}{3}$ & $ \frac{2}{3}$ & $1$ & $-1$  \\ \midrule 
$U(1)_R$ & 1 & 1 & 1 & 1 & 1 & 1 & 1 & 0 & 0 \\ 
$A_4$ & $\mathbf{1}$ & $\mathbf{1}$ & $\mathbf{1}$ & $\mathbf{1}$ & $\mathbf{1}$ & $\mathbf{1}$ & $\mathbf{3}$ & $\mathbf{1}$ & $\mathbf{1}$  \\ 
$Z_2$ & 0 & 0 & 0 & 1 & 1 & 1 & 1 & 1 & 0  \\ 
$Z_4$ & 0 & 2 & 1 & 0 & 0 & 3 & 1 & 0 & 0  \\ 
$Z_4$ & 1 & 2 & 1 & 1 & 0 & 3 & 0 & 0 & 0  \\ 
$Z_4$ & 1 & 1 & 0 & 3 & 1 & 0 & 0 & 0 & 0  \\ 
$Z_4$ & 1 & 2 & 1 & 1 & 0 & 3 & 1 & 0 & 0  \\ 
$Z_4$ & 0 & 0 & 1 & 3 & 1 & 2 & 0 & 1 & 1  \\
\bottomrule
\end{tabular}
\caption{\label{Tab:Matter+HiggsFields}
The matter and Higgs fields in our model and their quantum numbers.
}
\end{table}
Apart from the triplet flavons discussed above, we also have to introduce the singlet flavons given in Table \ref{Tab:FlavonFields}, which will all obtain a CP conserving real vev. The MSSM matter and Higgs fields transform under the additional flavour symmetries as indicated in Table \ref{Tab:Matter+HiggsFields}. Under $A_4$ only the right-handed down type quarks transform non-trivially (as a triplet).

We will use an effective operator description that should be viewed as the low-energy effective theory of the renormalizable model we will present in Section \ref{sec:Messenger}. This 'UV completion' of the model, together with the symmetries, determines which effective operators appear. In addition, having the  'UV completion' at hand increases the predictivity of the model (cf.~\cite{Varzielas:2010mp}) and allows to calculate the corrections from higher-dimensional operators to every desired order. This is particularly important here since one has to make sure that such operators do not induce a too large value of $\bar\theta$. 

After symmetry breaking, the mass matrices will be generated by the superpotential
\begin{align}
  \mathcal{W}_d &= Q_1 \bar d H_d \frac{\phi_2 \xi_d}{\Lambda^2} + Q_2 \bar d H_d \frac{\phi_1 \xi_d + \tilde \phi_2 \xi_s + \phi_3 \xi_t}{\Lambda^2}  +  Q_3 \bar d H_d \frac{\phi_3}{\Lambda} \;,\\
  \mathcal{W}_u &=  Q_1 \bar u_1 H_u \frac{\xi_u^2}{\Lambda^2}  +  Q_1 \bar u_2  H_u \frac{\xi_u \xi_c}{\Lambda^2}  +  Q_2 \bar u_2 H_u \frac{\xi_c}{\Lambda}   + (Q_2 \bar u_3 + Q_3 \bar u_2) H_u \frac{\xi_t}{\Lambda} + Q_3 \bar u_3 H_u\;,
\end{align}
which results from integrating out the heavy messenger fields. For the superpotential, we use a notation where prefactors 
are dropped for brevity, trivial $A_4$ contractions are not explicitly shown \footnote{The only non-trivial contraction is between $\bar d$ and the $\phi_i$, which form a singlet contracted by the $SO(3)$-type inner product '$\cdot$'.}
and where
$\Lambda$ denotes a generic messenger scale which is larger than the family
symmetry breaking scale $M_F$.
After plugging in the Higgs and flavon vevs we find the following mass matrices
\begin{align}
 M_d = \begin{pmatrix}
	0 & b_d & 0 \\
	b'_d & \ci c_d & d_d \\
	0 & 0 & e_d
       \end{pmatrix}
\quad\quad\text{and}\quad\quad
 M_u = \begin{pmatrix}
	a_u & b_u & 0 \\
	0 & c_u & d_u \\
	0 & d'_u & e_u
       \end{pmatrix} \;.
\end{align}
where we used the left-right convention $-\mathcal{L} = \overline{u^i_L} (M_u)_{ij} u^j_R + \overline{d^i_L}
(M_d)_{ij} d^j_R + \text{ H.c.}$. Note that due to the fundamental CP symmetry and its peculiar breaking pattern, eq.~\eqref{eq:CPvioaltingvev}, all entries are real apart from the
2-2 element of $M_d$. 
As discussed before, it predicts the right quark unitarity triangle\cite{Antusch:2009hq}
in terms of a phase sum rule
\begin{equation}
\alpha \approx \delta_{12}^d - \delta_{12}^u \approx 90^\circ  \;, 
\end{equation}
where the angle $\alpha$ of the CKM unitarity triangle
is measured to be close to $90^\circ$ \cite{Beringer:1900zz}. In this toy model, we concentrate on the explanation of CP violation in strong and weak interactions. Therefore, we are content with the prediction of the smallness of the strong CP phase and the correct CP phase in the CKM matrix. We are able to fit all masses and mixing angles (cf.~\cite{Antusch:2009hq}). A more realistic model should obviously aim at predicting the masses and mixing angles as well, which happens quite naturally in a GUT context, for instance. In fact, a similar texture has been obtained in a GUT based model \cite{Antusch:2013kna}, which might solve the strong CP problem as well. 

\subsection{Alignment}
\label{sec:Model:Flavon}

\begin{table}
\centering
\begin{tabular}{c ccccc ccccc cccc}
\toprule
 & $G_{\text{SM}}$ & $A_4$ & $U(1)_R$ & $Z_2$ & $Z_4$ & $Z_4$ & $Z_4$ & $Z_4$ & $Z_4$ \\ 
 \midrule 
$O_{1;3}$  & $(\mathbf{1},\mathbf{1},0)$ & $\mathbf{1}$  & 2 & 0 & 1 & 2 & 1 & 1 & 0\\
$O_{2;3}$  & $(\mathbf{1},\mathbf{1},0)$ & $\mathbf{1}$  & 2 & 0 & 3 & 1 & 1 & 0 & 0\\
$\tilde O_{1;2}$ & $(\mathbf{1},\mathbf{1},0)$ & $\mathbf{1}$  & 2 & 1 & 2 & 1 & 1 & 1 & 0\\
$\tilde O_{2;3}$ & $(\mathbf{1},\mathbf{1},0)$ & $\mathbf{1}$  & 2 & 1 & 1 & 1 & 0 & 0 & 0\\
\midrule
$A_{1}$ & $(\mathbf{1},\mathbf{1},0)$ & $\mathbf{3}$  & 2 & 0 & 2 & 2 & 2 & 2 & 0\\
$A_{2}$ & $(\mathbf{1},\mathbf{1},0)$ & $\mathbf{3}$  & 2 & 0 & 2 & 0 & 2 & 0 & 0\\
$A_{3}$ & $(\mathbf{1},\mathbf{1},0)$ & $\mathbf{3}$  & 2 & 0 & 0 & 2 & 0 & 0 & 0\\
$\tilde A_{2}$  & $(\mathbf{1},\mathbf{1},0)$ & $\mathbf{3}$  & 2 & 0 & 2 & 0 & 0 & 0 & 0\\
\midrule
$P$  & $(\mathbf{1},\mathbf{1},0)$ & $\mathbf{1}$  & 2 & 0 & 0 & 0 & 0 & 0 & 0\\
\bottomrule
\end{tabular}
\caption{\label{Tab:DrivingFields}
The driving field content of our model. Note that we only
show here one $P$ field. Indeed one has to introduce as
many $P$ fields as flavons to fix the phases of vevs. Since
they will have all the same quantum
numbers they will mix and we can go to a basis where
the terms to fix the phase for each flavon is separated
from the others. See the discussion in the appendix \ref{app:driving} for more details.
}
\end{table}

To obtain the vev structure given in Eq. (\ref{eq:CPvioaltingvev}) we make use of  the discrete vacuum alignment techniques mentioned in the strategy \Secref{sec:Strategy}. The resulting setup is rather simple. The symmetries of the model allows one to write down the potential ($A_i$, $\tilde A_2$ are $A_4$ triplets, $O_{i;j}$, $\tilde O_{i;j}$ and $P$ $A_4$ singlets)
\begin{align}\label{eq:W-alignment}
  \mathcal{W} =&\sum_{i=1}^{3}A_i \cdot (\phi_i \star \phi_i)  + O_{1;3} (\phi_1 \cdot \phi_3) + O_{2;3} (\phi_2 \cdot \phi_3)   + \frac{P}{\Lambda^{2}} \left( \phi_i^4 - M_F^4 \right) \\\nonumber
&+  \tilde A_2\cdot (\tilde \phi_2 \star \tilde \phi_2) + \tilde O_{1;2} (\phi_1 \cdot \tilde \phi_2) + \tilde O_{2;3} (\tilde\phi_2 \cdot \phi_3) + \frac{P}{\Lambda^{2}} \left( \tilde\phi_2^4 -M_F^4 \right)\;,
\end{align}
for the fields defined in table~\ref{Tab:FlavonFields} using the driving fields in table~\ref{Tab:DrivingFields}. We used the notation  '$\star$' ('$\times$') for the (anti-)symmetric triplet contraction of two triplets (see, for example, \cite{King:2006np}). We furthermore used a simplified notation where only one driving field $P$ is displayed. To fix the phases 
as in Eq. (\ref{eq:phaseswithZn}), one field for each operator is needed, as is reviewed in Appendix \ref{app:driving}.

 Note that the family symmetry breaking scale $M_F$ is real due to the underlying CP symmetry.

As has been discussed in Ref. \cite{Holthausen:2011vd}, highly symmetric vev configurations such as the one in \Eqref{eq:vevs}, may be interpreted as resulting from accidental symmetries of the flavon superpotential.  We will collectively denote these symmetries as $G_{\mathcal{W}}$. By calling them accidental we mean that they may be broken explicitly by other parts of the superpotential. The symmetry group $G_F$ of the full theory is in general only a subgroup $G_F\subset G_\mathcal{W}$. 

For the later discussion of corrections to the alignment it is useful to discuss the accidental symmetries of the flavon superpotential of Eq. (\ref{eq:W-alignment}) in some detail. First of all, we have the symmetries $Z_2^3$ given in Table \ref{Tab:acc-symm}, which have as a symmetric solution the vev alignment 
 \begin{equation}
  \langle \phi_1 \rangle \sim \text{e}^{\ci \alpha_1}\,\begin{pmatrix} 1 \\ 0 \\ 0 \end{pmatrix} \;,\quad
  \langle \phi_2 \rangle \sim \text{e}^{\ci \alpha_2}\,\begin{pmatrix} 0 \\ 1 \\ 0 \end{pmatrix} \;,\quad
  \langle \phi_3 \rangle \sim \text{e}^{\ci \alpha_3}\,\begin{pmatrix} 0 \\ 0 \\ 1 \end{pmatrix} \;,\quad
    \langle\tilde  \phi_2 \rangle \sim \text{e}^{\ci \tilde \alpha_2}\,\begin{pmatrix} 0 \\ 1 \\ 0 \end{pmatrix}.
 \end{equation}
 The phases of the vevs in \Eqref{eq:vevs}, $\alpha_i=0$ and $\tilde\alpha_2=\pi/2$, are a result of the CP transformation
 \begin{align}
 \tilde \phi_2 \rightarrow -\tilde \phi_2^* \;, \;\;\tilde{O}_{i;j}\rightarrow -\tilde{O}_{i;j}^* \;, \;\; \varphi\rightarrow \varphi^*  
 \label{eq:CP4}
 \end{align}
where $\varphi$ denotes all other fields in the theory.  All of these symmetries are not symmetries of the full theory but rather emerge as accidental symmetries of Eq.~(\ref{eq:W-alignment}) due to the chosen particle content and due to the symmetries of the original theory. 

Note that there are of course other discrete accidental symmetries whose symmetric solutions correspond to a vev configuration where, for instance, other fields have imaginary vevs. However, these solutions are physically distinct from our solution, as they correspond to different conserved subgroups~\cite{Michel:1980pc}. 
Since our alignment including phases is related to the accidental $Z_2^3$ and CP symmetry only correction terms which explicitly break one of these groups might disturb the structure of the vevs. We will show in \Secref{sec:Messenger} that the higher dimensional operators in our model indeed do not violate the accidental symmetries and hence there are no NLO corrections to the vev structure.

\begin{table}
\centering
\begin{tabular}{c cccc c cccc}
\toprule
 & $\phi_1$&$\phi_2$& $\phi_3$&$\tilde\phi_2$&$A_i$& $O_{1;3}$  &$O_{2;3}$  &$\tilde O_{1;2}$  &$\tilde O_{2;3}$  \\ 
 \midrule 
$Z_2$  & $S$ &$-S$ &$-S$ & $-S$&$S$  & - & + & - & +  \\
$Z_2$  & $-T^2ST$ &$T^2ST$ &$-T^2ST$ & $T^2ST$&$T^2ST$  & + & - & - & -  \\
$Z_2$  & $-TST^2$ &$-TST^2$ &$TST^2$ & $-TST^2$&$TST^2$  & - & - & + & -  \\
\bottomrule
\end{tabular}
\caption{\label{Tab:acc-symm}
Accidental symmetries of the flavon superpotential of Eq.~(\ref{eq:W-alignment}), which are left unbroken by the vev configuration. Note that this is not a symmetry of the full theory but rather emerges as a consequence of symmetries and particle content of the full theory. }
\end{table}

To see how the vacuum alignment follows dynamically from minimisation conditions in the supersymmetric limit we study the $F$-term condition 
\begin{align}
0=\frac{\partial \mathcal{W}}{\partial P}= \frac{1}{\Lambda^{2}} \left( \phi_i^4 - M_F^4 \right),
\label{eq:discrete-phase-freedom}
\end{align}
which forces the `fourth power of the flavon vevs to be real. For $\tilde \phi_2$ we choose the complex solution
while the other three $\phi_i$ flavon vevs are chosen
to be real.
The $F$-term conditions $0=\frac{\partial \mathcal{W}}{\partial A_i}$ force the the vev of the fields $\phi_1$, $\phi_2$, $\phi_3$ and $\tilde\phi_2$ to have at most one non-vanishing component while
$0=\frac{\partial \mathcal{W}}{\partial O_{i;j}}$ makes the vevs of the pairs $(\phi_1, \phi_3)$, $(\phi_2, \phi_3)$,  $(\phi_1, \tilde \phi_2)$ and  $(\tilde\phi_2, \phi_3)$ orthogonal. 
From these conditions the direction of $\langle\phi_2\rangle$ is not completely determined. There are two degenerate minima with $\langle \phi_2\rangle \sim (1,0,0)^T$ and
$\langle\phi_2\rangle \sim (0,1,0)^T$ (after we have chosen a basis where $\langle\phi_3\rangle \sim (0,0,1)^T$ and $\langle\phi_1\rangle \sim (1,0,0)^T$) and we choose the second one. 

The vev of the singlet flavons $\xi_i$ is determined via
\begin{equation}
 \mathcal{W} = \frac{P}{\Lambda^{2}} \left( \xi_i^4 - M_F^4 \right) + P \left( \xi_c^2 + \frac{\xi_c \xi_t^2}{\Lambda} - M_F^2 \right)  \;,
\end{equation}
where $i = d,s,u,t$. Note that an effective $\xi_c^4$ term
is allowed by the symmetries but not allowed by the
messenger sector which we will discuss in the next section.
We are working in a basis for
the $P$ fields in which the terms for $\xi_{d,s,u,t}$
are diagonal up to this order (Note that they do not mix with each
other, i.e.~no term $P \xi_u^2 \xi_t^2$ is allowed).
After these fields have received their (real) vev
also the phase of the $\xi_c$ flavon is fixed to be real.

Note that the UV completion discussed in the following paragraph allows additional higher-dimensional
operators, which are highly suppressed and do not change the alignment as discussed here. 

\subsection{The renormalizable superpotential and higher-dimensional Operators}
\label{sec:Messenger}

\begin{table}
\centering
\begin{tabular}{c ccc cccccccc}
\toprule
 & $SU(3)$ & $SU(2)$ & $U(1)_Y$ & $A_4$ & $U(1)_R$ & $Z_2$ & $Z_4$ & $Z_4$ & $Z_4$ & $Z_4$ & $Z_4$ \\ 
 \midrule 
$\Delta_1$, $\bar{\Delta}_1$ & $\mathbf{3}$, $\bar{\mathbf{3}}$  & $\mathbf{2}$, $\mathbf{2}$  & $ \frac{1}{3}$, $- \frac{1}{3}$ & $\mathbf{3}$, $\mathbf{3}$  & 1, 1 & 1, 1 & 3, 1 & 0, 0 & 0, 0 & 3, 1 & 3, 1 \\
$\Delta_2$, $\bar{\Delta}_2$ & $\mathbf{3}$, $\bar{\mathbf{3}}$  & $\mathbf{2}$, $\mathbf{2}$  & $ \frac{1}{3}$, $- \frac{1}{3}$ & $\mathbf{3}$, $\mathbf{3}$  & 1, 1 & 0, 0 & 3, 1 & 2, 2 & 1, 3 & 0, 0 & 2, 2 \\
$\Delta_3$, $\bar{\Delta}_3$ & $\mathbf{3}$, $\bar{\mathbf{3}}$  & $\mathbf{2}$, $\mathbf{2}$  & $ \frac{1}{3}$, $- \frac{1}{3}$ & $\mathbf{3}$, $\mathbf{3}$  & 1, 1 & 1, 1 & 3, 1 & 1, 3 & 0, 0 & 3, 1 & 2, 2 \\
\midrule
$\Upsilon_1$, $\bar{\Upsilon}_1$ & $\bar{\mathbf{3}}$, $\mathbf{3}$  & $\mathbf{2}$, $\mathbf{2}$  & $- \frac{1}{3}$ , $ \frac{1}{3}$  & $\mathbf{1}$, $\mathbf{1}$  & 1, 1 & 0, 0 & 3, 1 & 3, 1 & 0, 0 & 3, 1 & 3, 1 \\
$\Upsilon_2$, $\bar{\Upsilon}_2$ & $\bar{\mathbf{3}}$, $\mathbf{3}$  & $\mathbf{2}$, $\mathbf{2}$  & $- \frac{1}{3}$ , $ \frac{1}{3}$  & $\mathbf{1}$, $\mathbf{1}$  & 1, 1 & 0, 0 & 0, 0 & 0, 0 & 1, 3 & 0, 0 & 2, 2 \\
$\Upsilon_3$, $\bar{\Upsilon}_3$ & $\mathbf{3}$, $\bar{\mathbf{3}}$  & $\mathbf{1}$, $\mathbf{1}$  & $ \frac{4}{3}$, $- \frac{4}{3}$ & $\mathbf{1}$, $\mathbf{1}$  & 1, 1 & 1, 1 & 0, 0 & 1, 3 & 1, 3 & 1, 3 & 1, 3 \\
\midrule
$\Xi_1$, $\bar{\Xi}_1$ & $\mathbf{1}$, $\mathbf{1}$  & $\mathbf{1}$, $\mathbf{1}$  & 0, 0 & $\mathbf{1}$, $\mathbf{1}$  & 2, 0 & 0, 0 & 2, 2 & 2, 2 & 2, 2 & 2, 2 & 0, 0 \\
$\Xi_2$, $\bar{\Xi}_2$ & $\mathbf{1}$, $\mathbf{1}$  & $\mathbf{1}$, $\mathbf{1}$  & 0, 0 & $\mathbf{1}$, $\mathbf{1}$  & 2, 0 & 0, 0 & 2, 2 & 0, 0 & 2, 2 & 0, 0 & 0, 0 \\
$\Xi_3$, $\bar{\Xi}_3$ & $\mathbf{1}$, $\mathbf{1}$  & $\mathbf{1}$, $\mathbf{1}$  & 0, 0 & $\mathbf{1}$, $\mathbf{1}$  & 2, 0 & 0, 0 & 0, 0 & 2, 2 & 0, 0 & 0, 0 & 0, 0 \\
$\Xi_4$, $\bar{\Xi}_4$ & $\mathbf{1}$, $\mathbf{1}$  & $\mathbf{1}$, $\mathbf{1}$  & 0, 0 & $\mathbf{1}$, $\mathbf{1}$  & 2, 0 & 0, 0 & 2, 2 & 0, 0 & 0, 0 & 0, 0 & 0, 0 \\
\midrule
$\Sigma_1$, $\bar{\Sigma}_1$ & $\mathbf{1}$, $\mathbf{1}$  & $\mathbf{1}$, $\mathbf{1}$  & 0, 0 & $\mathbf{1}$, $\mathbf{1}$  & 2, 0 & 0, 0 & 0, 0 & 2, 2 & 0, 0 & 0, 0 & 2, 2 \\
$\Sigma_2$, $\bar{\Sigma}_2$ & $\mathbf{1}$, $\mathbf{1}$  & $\mathbf{1}$, $\mathbf{1}$  & 0, 0 & $\mathbf{1}$, $\mathbf{1}$  & 2, 0 & 0, 0 & 0, 0 & 0, 0 & 2, 2 & 2, 2 & 2, 2 \\
$\Sigma_3$, $\bar{\Sigma}_3$ & $\mathbf{1}$, $\mathbf{1}$  & $\mathbf{1}$, $\mathbf{1}$  & 0, 0 & $\mathbf{1}$, $\mathbf{1}$  & 2, 0 & 0, 0 & 0, 0 & 2, 2 & 0, 0 & 2, 2 & 0, 0 \\
$\Sigma_4$, $\bar{\Sigma}_4$ & $\mathbf{1}$, $\mathbf{1}$  & $\mathbf{1}$, $\mathbf{1}$  & 0, 0 & $\mathbf{1}$, $\mathbf{1}$  & 2, 0 & 0, 0 & 2, 2 & 2, 2 & 2, 2 & 2, 2 & 2, 2 \\
$\Sigma_5$, $\bar{\Sigma}_5$ & $\mathbf{1}$, $\mathbf{1}$  & $\mathbf{1}$, $\mathbf{1}$  & 0, 0 & $\mathbf{1}$, $\mathbf{1}$  & 2, 0 & 0, 0 & 0, 0 & 1, 3 & 2, 2 & 1, 3 & 2, 2 \\ 
\bottomrule
\end{tabular}
\caption{\label{Tab:MessengerFields}
The messenger field content of our model. Every line represents a messenger pair which receives a mass larger than the flavor breaking scale. In the main text we labelled the messenger mass scale generically with $\Lambda$.}
\end{table}

In this section we present an UV completion
of our toy model from \Secref{sec:Model} which justifies
completely the effective operators we have given there. We will
furthermore discuss all higher-dimensional operators which give
corrections to the mass matrices and to the flavon alignment.  We will
show that they do not alter the structure of the mass matrices and hence
our conclusions remain unchanged.

First of all, let us note that in the renormalizable superpotential
only one monomial term, $P M^2$, appears. All other fields cannot
appear alone due to the symmetries.

The messenger fields listed in table~\ref{Tab:MessengerFields}
receive pairwise a mass term
\begin{equation}
 \mathcal{W}_{\Lambda} = \sum_i M_{\Delta_i} \Delta_i \bar \Delta_i
	+ \sum_i M_{\Upsilon_i} \Upsilon_i \bar \Upsilon_i
	+ \sum_i M_{\Xi_i} \Xi_i \bar \Xi_i
	+ \sum_i M_{\Sigma_i} \Sigma_i \bar \Sigma_i \;,
\end{equation}
where we assume that all these masses are larger than the family
symmetry breaking scale $M_F$ and we have labelled them before
generically as $\Lambda$. Apart from these mass terms
also the combinations $Q_3  \Upsilon_1$ and $\xi_c \Sigma_4$
are allowed to have mass terms. But in fact we can rotate $Q_3$ and
$\Upsilon_1$ and $\xi_c$ and $\Sigma_4$ respectively so that these
combinations are massless.

\begin{figure}
\centering
\includegraphics[width=0.8\textwidth]{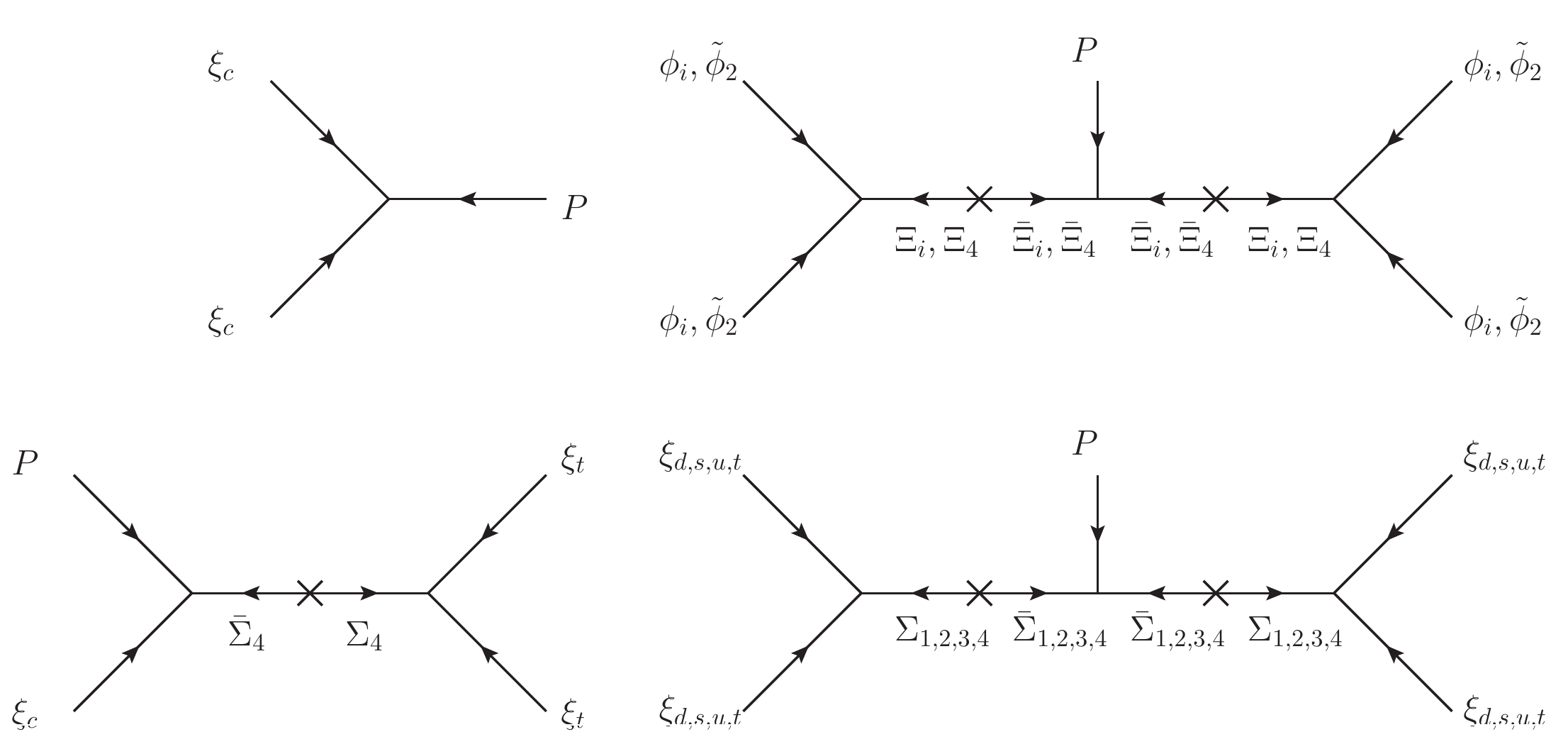}
\caption{
The supergraphs before integrating out the messengers for the flavon sector (only diagrams are shown which fix the phases of the flavon vevs).
\label{Fig:FlavonMessenger} }
\end{figure}

We come now to the trilinear couplings in the renormalisable
superpotential and start with the couplings involving only flavon
and  $\Xi$ and $\Sigma$ messenger fields
\begin{align}
 \mathcal{W}^{\text{ren}}_{\text{flavon}} &= P \xi_c^2 + P \xi_c \bar \Sigma_4 + \sum_{i=1}^4 P \bar \Xi_i^2 + \sum_{i=1}^4 P \bar \Sigma_i^2 + \sum_{i=1}^3 \Xi_i \phi_i^2 + \Xi_4 \tilde \phi_2^2  + \xi_d^2 \Sigma_1 + \xi_s^2 \Sigma_2 + \xi_u^2 \Sigma_3 \nonumber\\
& + \xi_t^2 \Sigma_4 + \sum_{i=1}^3 A_i \phi_i^2 + \tilde A_2 \tilde \phi_2^2 + O_{1;3} \phi_1 \phi_3 + O_{2;3} \phi_2 \phi_3 + \tilde O_{1;2} \phi_1 \tilde \phi_2 + \tilde O_{2;3} \tilde \phi_2 \phi_3 \nonumber\\
& + \xi_u \bar \Sigma_4 \Sigma_5 + \bar \Xi_1 \Xi_2 \bar \Sigma_3 + \Xi_1 \bar \Xi_2 \bar \Sigma_3 + \bar \Xi_1 \bar \Xi_2 \Sigma_3 + \Sigma_3 \bar \Sigma_5^2 \;.
\end{align}
After integrating out the messenger fields (as indicated in Fig.~\ref{Fig:FlavonMessenger}) we end up with the superpotential as given in \Secref{sec:Model:Flavon} plus  the following higher-dimensional operators:
\begin{align}
 \mathcal{W}^{\text{corr}}_{\text{flavon}} & = \frac{P}{\Lambda^4} ( \phi_1^2 \phi_2^2 \xi_u^2 + \xi_u^4 \xi_c^2 )
 + \frac{P}{\Lambda^5} \xi_c \xi_t^2 \xi_u^4
 + \frac{P}{\Lambda^6} ( \xi_t^4 \xi_u^4 + \xi_c^2 \xi_u^2 \phi_1^2 \phi_2^2 + \phi_1^4 \phi_2^4 + \xi_u^4 (\phi_1^4 + \phi_2^4)) \nonumber\\
 &
 + \frac{P}{\Lambda^7} \xi_c \xi_t^2 \xi_u^2 \phi_1^2 \phi_2^2
 + \frac{P}{\Lambda^8} \xi_u^2 (\xi_c^2 \xi_u^2 (\phi_1^4 + \phi_2^4) + \phi_1^2 \phi_2^2 (\xi_t^4 + \phi_1^4 + \phi_2^4)) \nonumber\\
 &
 + \frac{P}{\Lambda^9} \xi_c \xi_t^2 \xi_u^4 (\phi_1^4 + \phi_2^4)
 + \frac{P}{\Lambda^{10}} (\phi_1^4 + \phi_2^4) (\xi_t^4 \xi_u^4 + \xi_c^2 \xi_u^2 \phi_1^2 \phi_2^2 + \phi_1^4 \phi_2^4) \nonumber\\
 &
 + \frac{P}{\Lambda^{11}} \xi_c \xi_t^2 \xi_u^2 \phi_1^2 \phi_2^2 (\phi_1^4 + \phi_2^4)
 + \frac{P}{\Lambda^{12}} \xi_t^4 \xi_u^2 \phi_1^2 \phi_2^2 (\phi_1^4 + \phi_2^4)
\;.
\end{align}
Below Eq. (\ref{eq:W-alignment}) we had already inferred the fact that our vev alignment is a solution by 
looking at the accidental symmetries of the potential. We can repeat the same analysis here. It can easily be checked that as the messengers only produce operators where the fields $\phi_i$ appear squared, the relevant accidental symmetry $Z_2^3$ of Table~\ref{Tab:acc-symm} that enforces the vev directions is left-unbroken by the additional higher-dimensional operators. There are thus no corrections to the leading-order vev structure (their magnitude might of course be slightly corrected). Also the accidental CP transformation of Eq.~(\ref{eq:CP4}) is being left unbroken and therefore the phases of the vevs are not corrected. 

The same result can be obtained in the following way: 
First of all, note that $\tilde \phi_2$, $\phi_3$, $\xi_d$ and $\xi_s$
do not appear in these operators such that their phase is not
corrected. Also note that these operators only concern the
phases and not the directions of the triplet flavon vevs. For the remaining five flavons we cannot find a basis for the $P$
fields such that the phases are easily to be read off and in fact
the polynomial in these five fields is quite complicated. But still
it is easy to convince oneself that these flavon vevs remaining real is a viable
solution to the $F$-term conditions. Remember that all couplings are
real and hence the reality of the solutions is just a question of
having the right signs and moduli for the couplings.

\begin{figure}
\centering
\includegraphics[width=0.8\textwidth]{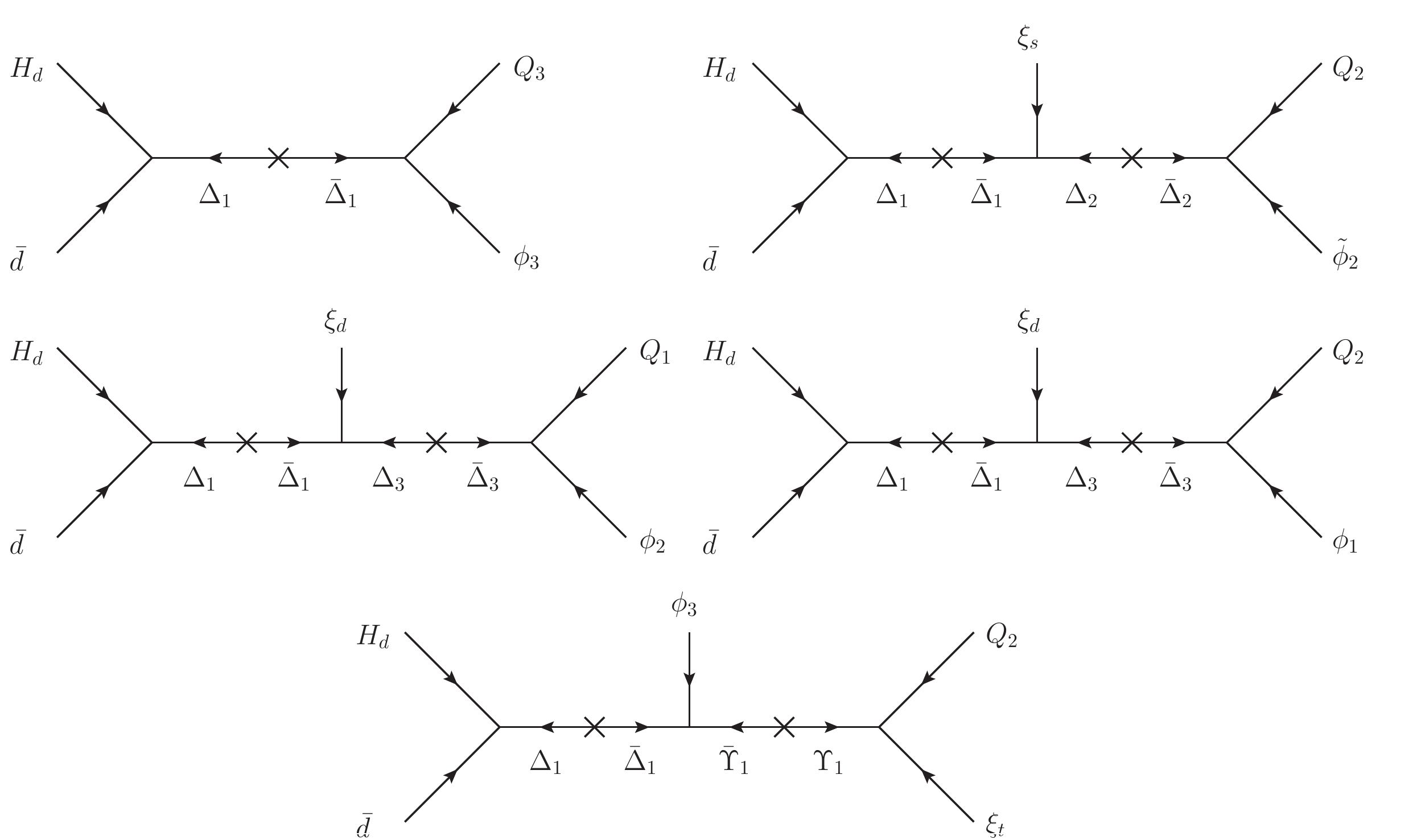}
\caption{
The supergraphs before integrating out the messengers for the down-type quark sector.
\label{Fig:DownMessenger} }
\end{figure}

We turn now to the down-type quark sector. Here the renormalisable
superpotential reads
\begin{align}
 \mathcal{W}^{\text{ren}}_{d} &= H_d \bar d \Delta_1 + Q_3 \phi_3 \bar \Delta_1 + \xi_s \bar \Delta_1 \Delta_2 + Q_2 \tilde \phi_2 \bar \Delta_2 + \xi_d \bar \Delta_1 \Delta_3 + Q_2 \phi_1 \bar \Delta_3 + Q_1 \phi_2 \bar \Delta_3 + \phi_3 \bar \Upsilon_1 \bar \Delta_1 \;,
\end{align}
which gives the effective operators as discussed in
\Secref{sec:Model:Matter}. We did not find any
higher-dimensional operators produced at tree-level that would contribute to the down type quark sector.

\begin{figure}
\centering
\includegraphics[width=0.8\textwidth]{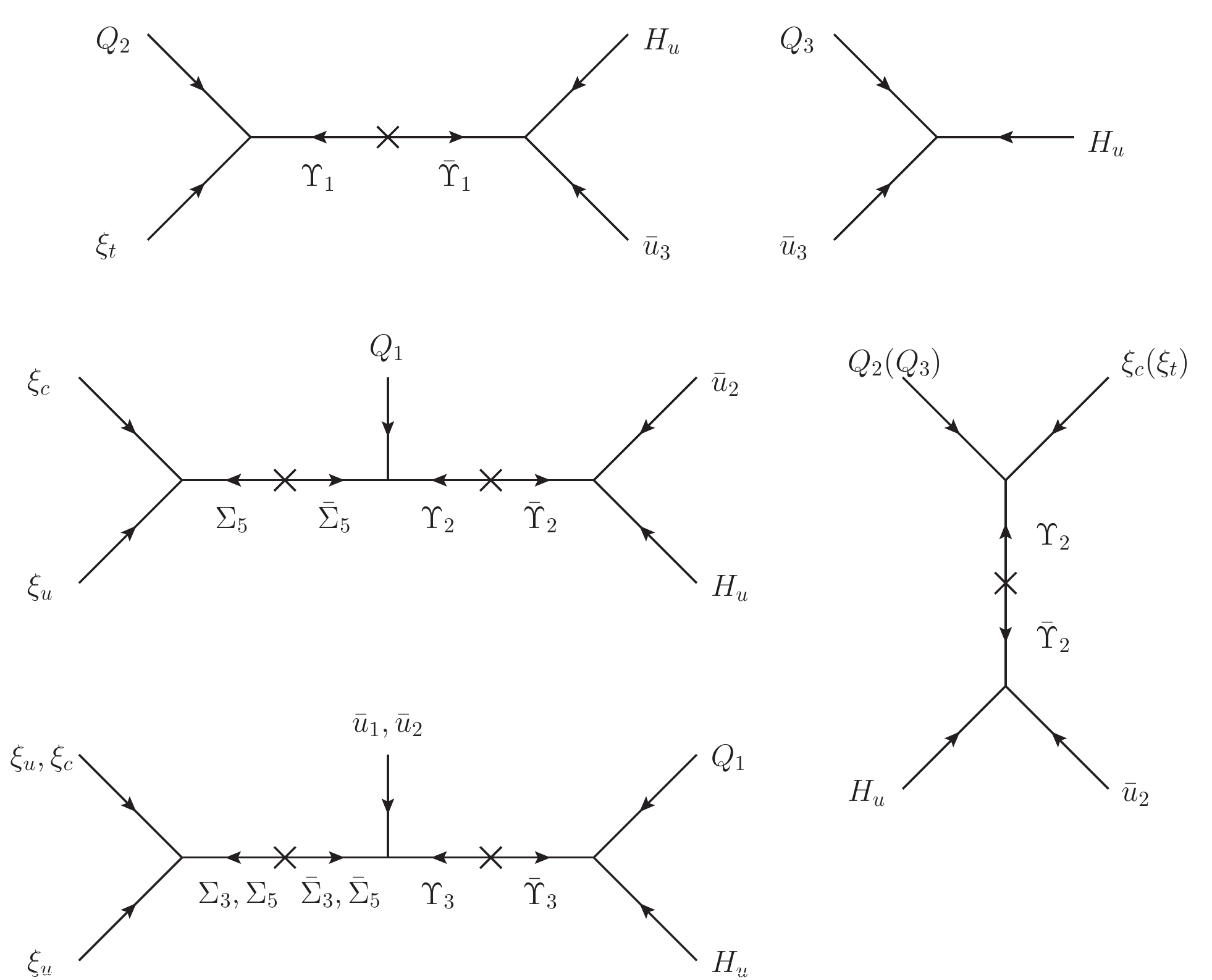}
\caption{ The supergraphs before integrating out the messengers for
the up-type quark sector.
\label{Fig:UpMessenger} }
\end{figure}

The last sector we are going to discuss here is the
up-type quark sector. Here the renormalisable
superpotential reads
\begin{align}
 \mathcal{W}^{\text{ren}}_{u} &= H_u Q_3 \bar u_3 + H_u \bar u_3 \bar \Upsilon_1 + Q_2 \xi_t \Upsilon_1 + H_u \bar u_2 \bar \Upsilon_2 + Q_2 \xi_c \Upsilon_2 + \xi_c \xi_u \Sigma_5 \nonumber\\
 & + Q_1 \bar \Sigma_5 \Upsilon_2  + \bar u_1 \bar \Sigma_3 \Upsilon_3 + H_u Q_1 \bar \Upsilon_3 + \bar u_2 \bar \Sigma_5 \Upsilon_3 + Q_3 \xi_t \Upsilon_2  + Q_2 \bar \Sigma_4 \Upsilon_2 + \xi_t \bar \Upsilon_1 \Upsilon_2 \;,
\end{align}
where again after integrating out the messengers we
get the effective operators as in \Secref{sec:Model:Matter},
see also Fig.~\ref{Fig:UpMessenger}. In contrast to the
down-type quark sector here are some additional
operators allowed which give (real) corrections to the
entries of the up-type quark mass matrix:
\begin{align}\label{eq:Wu-corr}
 \mathcal{W}^{\text{corr}}_{\text{u}} & = Q_1 \bar u_1 H_u \left( \frac{\xi_c^2 \xi_u^2 + \phi_1^2 \phi_2^2}{\Lambda^4} + \frac{\xi_c \xi_t^2 \xi_u^2}{\Lambda^5} + \frac{\xi_t^4 \xi_u^2}{\Lambda^6} \right)
+ Q_1 \bar u_2 H_u \frac{\xi_t^2 \xi_u}{\Lambda^3} + Q_2 \bar u_2 H_u \frac{\xi_t^2}{\Lambda^2} \;.
\end{align} 
These corrections are subleading real corrections to
real entries of the Yukawa matrix and hence do not
alter the fact that $\bar \theta=0$.

Finally, let us briefly comment on the effects anomalies might have on our results. The gauge symmetries remain anomaly free (after adding the leptons), because we do not add new chiral fermions, which are charged under the Standard Model gauge group. 
In addition, as we do not introduce non-trivial singlet representations of $A_4$, the $A_4$ group is anomaly free, but some of the auxiliary $Z_n$ symmetries appear to be anomalous.\footnote{For a general discussion of anomalies of discrete symmetry groups the reader is referred  to\cite{Araki:2007zza,Ishimori:2012zz}.}
However, since we do not specify here a complete model (including leptons, a SUSY Breaking Sector etc.), we cannot make definite statements about anomalies but we assume that the effects of anomalies are either cancelled in the complete theory or sufficiently small.

\section{Relation to other models based on spontaneous CP violation}
\label{sec:NelsonBarr}

In this section we want to discuss how our
class of models is related to other models explaining the smallness of the strong CP phase by a spontaneous breaking of CP. We will especially focus on the Nelson-Barr
models of spontaneous CP violation \cite{Nelson:1983zb,
Barr:1984qx} being the first and most studied models. Although there are certain similarities, our model,
for instance, does not fulfil the Barr criteria \cite{Barr:1984qx}:

He classified the fields in two sets, the low energy fermions $F$ and heavy vector-like fermions $R=C+\bar C$. Then, there are two sufficient conditions for a solution of the strong CP problem:
\begin{enumerate}
\item At the tree level there are no Yukawa or mass terms
 coupling $F$ fermions to $\bar C$ fermions, or $C$ fermions
 to $C$ fermions.
\item The CP-violating phases appear at the tree
 level only in those Yukawa terms that couple $F$ fermions
 to $R =  C + \bar C$ fermions.
\end{enumerate}
In such a setup the determinant of the mass matrices
is real and the anomalous contribution to $\bar \theta$
hence vanishes. 

In our class of models, the $F$ fields are the ordinary quark
fields and the $C + \bar C$ fields are the heavy messenger
fields, see \Secref{sec:Messenger}. As all messengers carry a non-vanishing $U(1)_Y$ charge, there are no $C C$ mass terms. Furthermore since the only allowed
Yukawa coupling on a renormalisable level is the (real) top Yukawa coupling, condition 2 is fulfilled as well. However, there are couplings of the $F$ fields to $C$ as well as $\bar C$ and therefore condition 1 is not satisfied.

Let us now have a closer look at the mass matrices in the full theory following Ref.~\cite{Barr:1984qx} to understand why the strong CP phase vanishes.
It is straightforward to see that the determinant of the up-type quark mass matrix is real, since all couplings are real due to the imposed CP symmetry and the only flavon with a complex vev, $\tilde \phi_2$, does not couple to the up sector. 
Therefore, it remains to study the mass matrix $q^c M q$ in the down-type sector, where in our case
\begin{equation}
\begin{split}
 q &= ((d_L)_i, (\Delta_1)_i, (\Delta_2)_i, (\Delta_3)_i, \bar \Upsilon_1, \bar \Upsilon_2), \\
 q^c &= ((\bar d_R)_1,(\bar d_R)_2,(\bar d_R)_3, (\bar \Delta_1)_i, (\bar \Delta_2)_i, (\bar \Delta_3)_i, \Upsilon_1, \Upsilon_2)
 \end{split}
\end{equation}
with flavour index $i=1,2,3$. Note that we have labelled here the components of the messengers with the same $SU(3) \times U(1)_\text{em}$
as the down-type quarks with the same letter as the fields before symmetry breaking for the sake of simplicity.
We find then
\begin{equation}
 M \sim \begin{pmatrix}
      0 & 0 & 0 & \vev{\phi_2}^T & 0 & 0 \\
      0 & 0 & \vev{\tilde \phi_2}^T & \vev{\phi_1}^T & \vev{\xi_t} & \vev{\xi_c} \\
      0 & \vev{\phi_3}^T & 0 & 0 & 0 & \vev{\xi_t} \\
      \vev{H_d} & M_{\Delta_1} & 0 & 0 & 0 & 0 \\
      0 & \vev{\xi_s} & M_{\Delta_2} & 0 & 0 & 0 \\
      0 & \vev{\xi_d} & 0 & M_{\Delta_3} & 0 & 0 \\
      0 & \vev{\phi_3}^T & 0 & 0 & M_{\Upsilon_1} & \vev{\xi_t} \\
      0 & 0 & 0 & 0 & 0 & M_{\Upsilon_2} \\
     \end{pmatrix} \;,
\end{equation}
where we have dropped order one coefficients
 and note that the matrix is not quadratic because
we have not expanded all flavour indices.

In the Nelson-Barr models the entries above the messenger
masses or left of the messenger masses vanish, which is not satisfied in our model. And in fact, calculating the
determinant naively and using arbitrary complex vevs
for the flavons the determinant would be complex and
the strong CP problem would not be solved.
However, inserting the alignment of the flavon vevs we find for the determinant
\begin{equation}\label{eq:detMd}
 \det M \sim \vev{H_d}^3 M_{\Delta_2}^3 M_{\Delta_3}^3 M_{\Upsilon_1} M_{\Upsilon_2} \langle \xi_d^2 \rangle \langle \phi_1 \rangle \langle \phi_2 \rangle \langle \phi_3 \rangle \,
\end{equation}
where we have again dropped (real) order one coefficients.
All flavon vevs in \Eqref{eq:detMd} are real due the flavon vev alignment mechanism and hence the determinant is real. Together with the real determinant of the up-type quark mass matrix, the total anomalous
correction of the quarks and heavy messengers to $\bar \theta$ vanishes.

The main difference to other models based on spontaneous breaking of CP as solution are the textures of the quark mass matrices. For example, the models in Ref.~\cite{Barr:1996wx,Glashow:2001yz} rely on triangular quark mass matrices. As long as the vevs of the Higgs coupling to the diagonal do not receive a complex vev, the determinants are real and hence $\bar\theta=0$. The solution proposed in Ref.~\cite{Masiero:1998yi} by Masiero and Yanagida relies on the generation of Hermitian quark mass matrices using flavon fields in the adjoint representation of the family symmetry $SU(3)_F$. This is similar to the solution in left-right symmetric models.

\section{Summary and conclusions}
\label{sec:conclusions}
In this work we have studied a novel approach to solve the
strong CP problem in the context of spontaneous CP violation
without the need for an axion.
We assume CP to be a fundamental symmetry of nature and use
discrete, Abelian and non-Abelian (family) symmetries to
break it in such a way that the anomalous contribution to
the CP violating QCD parameter $\bar \theta$ from the quark
mass matrices vanishes at tree-level. Simultaneously
the CKM phase is predicted to have its observed large
value in a natural way.

An essential ingredient of this approach is that the
phases of the symmetry breaking vevs are fixed to
certain discrete values with either being real or
purely imaginary in the simplest possible setup. In our toy
model we have used for this purpose the discrete
vacuum alignment method proposed in \cite{Antusch:2011sx}.
This method is based on supersymmetry and hence the class
of models which we propose here is supersymmetric
although in principle a non-supersymmetric version
reproducing the texture from eq.~\eqref{eq:MdStructure}
could do the same trick.

But in fact, supersymmetry does not only help to
fix the flavon vev phases but it also forbids via
the non-renormalisation theorem the appearance of
new, unwanted operators in the superpotential
from loop corrections, which could spoil our solution
for the strong CP problem. In fact, as long as SUSY remains
unbroken $\bar \theta$ remains zero on the perturbative level.
Furthermore, if the SUSY breaking sector itself conserves CP
and respects the discrete (family) symmetries the strong CP
phase can still be expected to be small enough to be in
agreement with experimental data. This remains true even after inclusion of
supergravity effects which are known to possibly cause a
sizeable misalignment for the trilinear SUSY
breaking couplings compared to the Yukawa couplings.

To demonstrate that our suggested class of models is in
principle realistic, we present an explicit supersymmetric
toy flavour model for the quark sector based on the family
symmetry $A_4$ with an $U(1)_R$ symmetry and the shaping
symmetry $Z_2 \times Z_4^5$. The shaping symmetry does not
only forbid unwanted operators in our superpotential but
also provides a mechanism to fix the phases of
the flavon vevs via the discrete
vacuum alignment method. We present an UV completion
of the model in that sense that we give a list of heavy messenger
fields which generate the desired effective operators after
being integrated out. This UV completion is very predictive
because it defines in combination with the symmetries the
set of allowed operators up to an arbitrary mass dimension.
Hence, we can show explicitly that our solution for the
strong CP problem is not affected by higher order corrections
(ignoring non-perturbative and SUSY breaking effects).

Finally, we discuss the relation between our novel class
of models to the well known Nelson-Barr models
\cite{Nelson:1983zb, Barr:1984qx} as well as other models which also solve the strong
CP problem in terms of spontaneous CP violation.
In the Nelson-Barr models direct couplings between the light sector and the heavy
sector are partially forbidden in such a way that the total mass
matrix exhibits a special block structure. This is different in our class of models, where all light fields can couple to all heavy
messenger fields in principle. The determinant of the total mass
matrix in their case is real due to the mentioned block structure,
while in our case it is real due to our vacuum alignment (including
phases).

The class of models presented here casts new light on an old
problem, the strong CP problem. 
There have been several previous attempts to solve it in terms of spontaneous CP violation in combination with flavour symmetries but our strategy differs significantly from the other approaches.
Most notably, we simultaneously have large CP violation in the CKM matrix with a right-angled unitarity triangle in a natural way,
without any contribution to $\bar \theta$ from the quark mass matrices.
Furthermore, the techniques to handle the symmetry breaking of
discrete non-Abelian family symmetries, like in our example model
$A_4$, was first developed in the context of the large leptonic
mixing angles and finds here an unexpected new application. Also
the method to fix the flavon vev phases was developed to give a
dynamical explanation for the phase sum rule but was then in
succeeding papers used in the lepton sector as well. Of course,
as a next step, it will be interesting, for instance,  to study the embedding of this mechanism in a GUT context
and the consequences for cosmology, e.g.\ for baryogenesis via the Leptogenesis
mechanism.

\section*{Acknowledgements}
The work of M.~Spinrath was supported by the ERC Advanced Grant no. 267985 ``DaMESyFla'', by the EU Marie Curie ITN ``UNILHC'' (PITN-GA-2009-237920) and the European Union FP7 ITN invisibles (Marie Curie Actions, PITN-GA-2011-289442-INVISIBLES). The work of S.~Antusch was supported by the Swiss National Science Foundation. This work was supported in part by the Australian Research Council.

\appendix
\section{The $\mathbf{P}$ driving fields}
\label{app:driving}
To fix the phases of the flavon vevs we have introduced a generic driving
field $P$.  But in fact, we need as many copies of $P$ as phases we want
to fix. In the appendix of \cite{Antusch:2011sx}, it was explicitly shown for two
flavon and $P$ fields how this form can be achieved after a suitable
basis choice for the $P$ fields. Here we want to extend the discussion to the case of an arbitrary number of fields.

Say we have a theory with $m$ flavon fields $\{\phi_1, \dots, \phi_m   \}$ out of which we can form $n$ ($m\leq n$) total singlet operators $\mathcal{O}_i=\prod _{l=1}^{m}\phi_{l}^{k^{(i)}_l}$ of order $\sum_{l=1}^{m} k_l^{(i)} \leq N$ and $k_l^{(i)}\in \{0,1, \dots, N \}$. We can then introduce (at least) $n$ fields $P_i$ with R-charge $2$\footnote{If we introduce more fields, one can always go to a basis where only the first $n$ couple to the operators $\mathcal{O}_i$. }. The $U(1)_R$ symmetric superpotential involving the $P_i$ fields is given by
\begin{align}
\mathcal{W}=P_i (\alpha_{ij} \mathcal{O}_j+m_i),
\end{align} 
where $\alpha\equiv (\alpha_{ij})$ is a real (because of CP) and invertible\footnote{A singular matrix would imply fine-tuning and can therefore be discarded.} matrix and $\vec{m}$ is real (again, because of CP) with either sign. The $n$ F-term conditions 
\begin{align}
\frac{\partial}{\partial P_i}\mathcal{W}=\alpha_{ij} \mathcal{O}_j+m_i =0
\end{align} 
give the  $n$ conditions
$
\alpha_{ij}\,\mathrm{Im} \mathcal{O}_j =0
$
 on the phases. Invertability of $(\alpha)_{ij}$ implies $\mathrm{Im} \mathcal{O}_j =0$ for all $j$, or 
 \begin{align}
k^{(i)}_j \mathrm{arg}  \phi_j =0 \mod{\pi}.
 \end{align}
Note that the $k^{(i)}_j$ are integer valued and bounded by $N$, the maximal considered operator dimension. To determine all phases of the $\phi_i$ the number of operators $n$ should be larger or equal to $m$, the number of fields. 

 In any case, the preceding discussion shows that one can always introduce fields $P$ such that the conditions $\mathrm{Im} \mathcal{O}_i=0$ are enforced. A concrete implementation of the P sector is therefore not necessary as long as the operator $\mathcal{O}_i$ set is specified and our discussion in the main part of the text goes through. In our case the operator set consists of (to leading order) $m$ operators of the form
 $\mathcal{O}_i = \phi_{i}^{k_i}$, $i= 1,\ldots,m$. Because of this construction, we can predict non-trivial phases.

The very same arguments apply for the fields $\tilde{\phi}_2$, $\phi_3$
and $\xi_d$ because there are no higher-dimensional operators correcting
their phases. We can decouple them from the other flavons in the above
mentioned way and we find the quoted phases for them.

It is slightly more complicated for the other fields because the
higher-dimensional operators introduce some mixing between the flavons and it is not possible to go to the convenient basis for the $P$ fields.
However the additional terms cannot alter our alignment and phases because they preserve the accidental
$Z_2^3$ and CP symmetry of the superpotential without higher order
corrections, as we discussed in section \ref{sec:Messenger}. Alternatively one could expand the vevs around their real leading order values and show that they remain real to all orders by choosing appropriate signs and
moduli of the couplings of the higher-dimensional operators.

\end{document}